\documentclass[twocolumn,showpacs,preprintnumbers,amsmath,amssymb]{revtex4}

\usepackage{graphicx}
\usepackage{dcolumn}
\usepackage{bm}
\usepackage[utf8]{inputenc}

\begin{document}

\newcommand{\bs}[1]{{\boldsymbol #1}}
\newcommand{\mb}[1]{{\mathbf #1}}
\renewcommand{\l}{\left}
\renewcommand{\r}{\right}
\newcommand{\mi}{\mathrm i}
\newcommand{\me}{\mathrm e}
\newcommand{\md}{\mathrm d}
\newcommand{\KM}[1]{D^{(#1)}}
\newcommand{\6}[2]{\frac{\partial #1}{\partial #2}}


\title{Estimation of Kramers-Moyal coefficients at low sampling rates}

\author{Christoph Honisch}
 \email{c.honisch@uni-muenster.de}
\author{Rudolf Friedrich}%
\affiliation{%
Institute for Theoretical Physics, University of Muenster, D-48149 Muenster, Germany
}%

\date{\today}

\begin{abstract}
A new optimization procedure for the estimation of Kramers-Moyal coefficients from stationary, one-dimensional, Markovian time series data is presented.
The method takes advantage of a recently reported approach that allows to calculate exact 
finite sampling interval effects by solving the adjoint Fokker-Planck equation. Therefore it is well suited for the analysis of sparsely sampled time series. The optimization can be performed either making a parametric ansatz for drift and diffusion functions or also parameter free. We demonstrate the power of the method in several numerical examples with synthetic time series.
\end{abstract}

\maketitle

\section{Introduction}
The behavior of complex systems consisting of a large number of degrees of freedom can
often be described by low dimensional macroscopic order parameter equations \cite{haken1}. Thereby the influence
of the microscopic degrees of freedom is treated via noise terms of Langevin type \cite{risken}. In case of a single
order parameter $q(t)$ its time evolution can be described by
\begin{equation} \label{eq:langevin}
  \dot q = h(q,t) + g(q,t)\Gamma(t)
\end{equation}
where $\Gamma(t)$ is a Gaussian distributed white noise term satisfying $\langle \Gamma(t) \rangle = 0$ and
$\langle \Gamma(t) \Gamma(t') \rangle = \delta(t-t')$. Here and in the following Ito's interpretation of
stochastic integrals is used \cite{risken}. 

The same information is contained in the corresponding Fokker-Planck equation (FPE) for the probability density function
of $f_q(x,t)$
\begin{equation}
  \frac{\partial f_q(x,t)}{\partial t} = \hat L(x,t) f_q(x,t)~.
\end{equation}
Here we have introduced the Fokker-Planck operator
\begin{equation}
  \hat L(x,t) = -\frac{\partial}{\partial x} \KM 1 (x,t) + \6{^2}{x^2} \KM 2 (x,t)
\end{equation}
which contains the Kramers-Moyal (KM) coefficients
\begin{equation} \label{eq:kmcoeff}
  \KM n (x,t) = \lim_{\tau \rightarrow 0} \frac{1}{n! \tau} \l.\l \langle [q(t+\tau)- q(t) ]^n  \r \rangle \r|_{q(t) = x}
\end{equation}
also referred to as drift and diffusion for $n=1$ and $n=2$, respectively. The connection to the functions $g$ and $h$
in Eq. \eqref{eq:langevin} is $h(x,t) = \KM 1 (x,t)$ and $g(x,t) = \sqrt{2 \KM 2(x,t)}$.

As was recently shown \cite{siegert98pla,friedrich2000pla}, it is possible to set up an equation of the form \eqref{eq:langevin} by estimating the conditional
averages in \eqref{eq:kmcoeff} from a data set of the variable $q(t)$. This method was applied in various fields of science, see Ref. \cite{friedrichEnc} for an overview.

There are two major problems connected to the estimation of drift and diffusion coefficients from measured ``real world'' time series. The first problem consists in the occurence of measurement noise. In Ref. \cite{kleinhans2007pre} it was shown that measurement noise spoils the Markov property, the latter being a requirement for the KM analysis. A promising approach to handle Gaussian distributed exponentially correlated measurement noise was recently proposed by Lehle \cite{lehle_pre}.

The other problem in the Kramers-Moyal analysis is that one has to perform the limit $\tau \rightarrow 0$, while data sets are recorded at
finite sampling intervals. Also in real world processes the intrinsic noise is not strictly $\delta$-correlated, which results in a finite Markov-Einstein time, i.\,e., a finite time interval $\tau_{\text{ME}}$ such that for time intervals $\tau<\tau_{\text{ME}}$ the Markov property does no
longer hold. It is observed that in case of a finite Markov-Einstein time, the KM coefficients go to zero with decreasing time interval $\tau$.

Ragwitz and Kantz \cite{ragwitz2001prl} were the first who presented a formular to estimate the KM coefficients that takes into account finite sampling interval effects at first order in the sampling interval.
In a comment on this article Friedrich \emph{et al.} \cite{friedrich2002prl} presented correction terms in form of an infinite series expansion in the sampling interval. Very recently Antenedo \emph{et al.} presented exact analytical expressions for the finite time KM coefficients (s. Eq. \eqref{eq:kmmn}) for processes with linear drift and quadratic diffusion \cite{antenedo2009pre} and later for other common processes \cite{antenedo2010pre}. 

A very elegant way to obtain finite time KM coefficients for arbitrary (but sufficiently smooth) drift and diffusion terms was recently presented by Lade \cite{lade2009pla}. He reinterpreted the series expansion presented in \cite{friedrich2002prl} in a way that finite time coefficients can be obtained by solving the adjoint Fokker-Planck equation. Since this can be done at least numerically, Lades method opens up the possibility to deduce the true KM coefficients from measured finite time coefficients by an optimization approach. This is the topic of the present work. 

Of course, finite time KM coefficients can also be obtained by simulating Langevin equations and measuring the conditional moments at a finite $\tau$. This was done in the iterative method developed by Kleinhans \emph{et al.} \cite{Kleinhans2005pla,kleinhans2007pla}. But since this is numerically very demanding, the method was only applied to situations where very few parameters had to be optimized. In this article we show that an optimization based on Lades method can even be performed without a parametric ansatz for drift and diffusion coefficients, which should make it more applicable for a larger class of diffusion processes.

In the next section we review the method of Lade \cite{lade2009pla} that allows for a calculation of exact finite time effects. The following section gives a description of our new optimization procedure. Section \ref{sec:exmpl} contains four numerical examples in which the functionality of our method is demonstrated.

\section{Exact finite sampling interval effects} \label{sec:fse}

From now on we assume the Langevin process of interest to be stationary, i.\,e., drift and diffusion do not explicitly depend on time.
We define the \emph{finite time coefficients} as
\begin{equation} \label{eq:kmmn}
  \KM n_{\tau}(x) = \frac{1}{n!\tau} M^{(n)}_{\tau}(x)
\end{equation}
with the conditional moments
\begin{equation} \label{eq:mn1}
  M^{(n)}_{\tau}(x) = \int_{-\infty}^{\infty} (x'-x)^n p(x',t+\tau|x,t) \md x'~.
\end{equation}
The conditional probability density function $p(x',t+\tau|x,t)$ is the solution of the corresponding FPE with the initial condition $\delta(x'-x)$, so it can be expressed as
\begin{equation}
  p(x',t+\tau|x,t) = \me^{\hat L(x')\tau} \delta(x'-x)~.
\end{equation}
Inserting this in \eqref{eq:mn1} results in
\begin{align}
  M^{(n)}_{\tau}(x) &= \langle (x'-x)^n | \me^{\hat L(x')\tau} | \delta(x'-x) \rangle  \nonumber\\
                      &= \langle \me^{\hat L^{\dag}(x')\tau} (x'-x)^n | \delta(x'-x) \rangle \nonumber \\
                      &= \l.\me^{\hat L^{\dag}(x')\tau} (x'-x)^n \r|_{x'=x}~, \label{eq:mn2}
\end{align}
where we use the notation $\langle f | g \rangle= \int_{-\infty}^{\infty} f(x') g(x') \md x'$ for the inner product. $\hat L^{\dag}$ is the adjoint Fokker-Planck operator
\begin{equation}
 \hat L^{\dag}(x') = \KM 1(x') \6 {}{x'} + \KM 2(x') \6{^2}{x'^2}~.
\end{equation}
The main point of Lade's article \cite{lade2009pla} is to interpret eq. \eqref{eq:mn2} as the solution to the partial differential equation
\begin{equation} \label{eq:AFP}
\l| \begin{aligned}
  \6{W_{n,x}(x',t)}{t} &= \hat L^{\dag}(x') W_{n,x}(x',t) \\
  W_{n,x}(x',0) &= (x' - x)^n
\end{aligned}  \r.
\end{equation}
at $t=\tau,~x=x_0$, i.\,e.:
\begin{equation} \label{eq:mnw}
  M^{(n)}_{\tau}(x) = W_{n,x}(x,\tau)~.
\end{equation}
For simple drift and diffusion coefficients, eq. \eqref{eq:AFP} can be solved analytically. E.\,g., for an Ornstein-Uhlenbeck process \cite{risken} with  $\KM 1(x) = -\gamma x$ and $\KM 2(x) = D$, one obtains \cite{lade2009pla}
\begin{align}
W_{1,x}(x',t) &= x'\me^{-\gamma t} - x \\
W_{2,x}(x',t) &= \l(x'\me^{-\gamma t} - x\r)^2 + \frac D{\gamma}\l(1-\me^{-2\gamma t}\r)~.
\end{align}
With Eq. \eqref{eq:kmmn} and \eqref{eq:mnw} we get
\begin{align}
  \KM 1_{\tau}(x) &= -\frac{x}{\tau} \l(1-\me^{-\gamma \tau}\r) \\
  \KM 2_{\tau}(x) &= \frac 1{2\tau}\l[ x^2\l(1-\me^{-\gamma \tau} \r)^2 + \frac D{\gamma} \l( 1-\me^{-2\gamma \tau} \r) \r]~.
\end{align}
A process with linear drift $\KM 1(x) = -\gamma x$ and quadratic diffusion $\KM2(x) = \alpha + \beta x^2$ gives the same finite time drift as for the Ornstein-Uhlenbeck process. For the diffusion we obtain
\begin{eqnarray}
  W_{2,x}(x',t) = x^2 - \frac{\alpha}{\beta-\gamma}\l(1-\me^{2(\beta-\gamma)t} \r) \nonumber \\
   + x'^2\me^{2(\beta-\gamma)t} - 2xx'\me^{-\gamma t}~,
\end{eqnarray}
which leads to
\begin{eqnarray}
  \KM 2_{\tau}(x) =  \frac 1{2\tau} \l[x^2 \l(1 + \me^{2(\beta-\gamma)\tau} - 2 \me^{-\gamma \tau} \r)\r. \nonumber \\ - \l.\frac{\alpha}{\beta-\gamma} \l( 1 - \me^{2(\beta-\gamma)\tau}\r) \r]~.
\end{eqnarray}
If an analytical solution cannot be obtained one has to solve eq. \eqref{eq:AFP} numerically up to $t=\tau$ for all $x$ values of interest.

An alternative way to calculate finite time effects would be to solve the real FPE, instead of the adjoint FPE, which yields the whole transition pdf. But this would involve a Dirac $\delta$-function as an initial condition which is expected to cause numerical problems. The adjoint FPE can be easily solved via a simple forward-time centered-space scheme. For the spatial derivatives on the left and right boundaries we use second order forward and backward finite differences, respectively.

\section{The optimization procedure}

\begin{figure}
  \footnotesize{\input{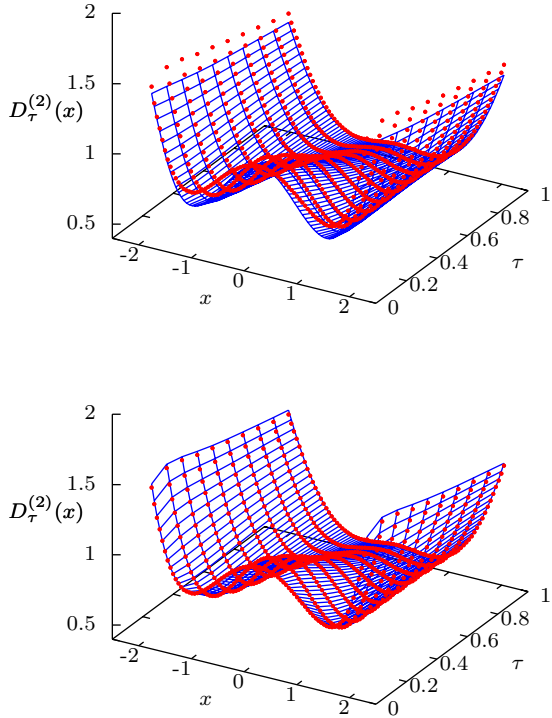}} 
  \caption{(Color online) Illustration of the optimization procedure. The red dots in both graphs show the estimated finite time coefficients $\hat{D}^{(2)}_{\tau}(x) =  \hat M^{(2)}_{\tau}(x)/(2\tau)$ for the example of Sec. \ref{sec:bistable}. The blue surface in the top panel corresponds to $\KM 2_{\tau}(x,\sigma_{\text{ini}})$. By minimizing $V(\sigma)$ we seek a set of parameters $\sigma$ such that $\KM{1,2}_{\tau}(x,\sigma)$ conforms to $\hat{D}^{(1,2)}_{\tau}(x)$, respectively, as is the case for the diffusion in the lower panel.} \label{fig:3d}
\end{figure}

The first step of the optimization is to estimate the conditional moments \eqref{eq:mn1} for a set of $\tau$ values $\{\tau_1,\dots,\tau_M\}$, $\tau_i < \tau_{i+1}$, and a set of $x$ values $\{x_1,\dots,x_N\}$, $x_i < x_{i+1}$. The latter should be the same values that are later on used for the numerical integration of the adjoint FPE. In a histogram based regression the size of the bins located at $x_i$ is limited through the available amount of data. Therefore a kernel based regression as described in \cite{lamouroux2009pla} is favorable which results in a smooth curve. We denote the estimated conditional moments by $\hat M^{(1,2)}_{\tau_i}(x_j)$. It is also important to calculate statistical errors $\hat \sigma_{ij}^{(1,2)}$.

The optimization can be performed with or without the use of parameterized drift and diffusion functions. In the former case one has to embed the drift and diffusion functions into a family of functions $\KM{1}(x,\sigma)$ and $\KM{2}(x,\sigma)$, respectively, with a set of parameters denoted by $\sigma$. 

In the latter case one has to define a set of sampling points $\{x^s_1,\dots,x^s_K \}$, $K<N$, and represent $\KM 1$ and $\KM 2$ as a spline interpolation through these sampling points. Then the set of parameters to be optimized is $\sigma=\{\KM 1(x^s_1),\dots,\KM 1(x^s_K),\KM 2(x^s_1),\dots,\KM 2(x^s_K) \}$. In both cases $\KM 1_{\tau_1}$ and $\KM 2_{\tau_1}$ can be used to construct an initial guess $\sigma_{\text{ini}}$.

For a specific set of parameters $\sigma$, the conditional moments \eqref{eq:mn1} can be calculated as described in sec. \ref{sec:fse}, yielding $M^{(1,2)}_{\tau_i}(x_j,\sigma)$. Since these computations are to be performed for each $x_j$ individually, it is very easy and efficient to parallelize this part for the use on parallel computers.

The final step is to find the minimum of the least square potential
\begin{eqnarray}
  V(\sigma) = \sum_{i=1}^M \sum_{j=1}^N \l[ \frac{\l\{ \hat M^{(1)}_{\tau_i}(x_j) -  M^{(1)}_{\tau_i}(x_j,\sigma) \r\}^2}{\l(\hat \sigma_{ij}^{(1)} \r)^2  } \r. \nonumber\\+ \l.\frac{\l\{ \hat M^{(2)}_{\tau_i}(x_j) -  M^{(2)}_{\tau_i}(x_j,\sigma) \r\}^2}{\l(\hat \sigma_{ij}^{(2)} \r)^2  } \r]~. \label{eq:pot}
\end{eqnarray}
Fig. \ref{fig:3d} illustrates the idea of this procedure. 

For the optimization we use a trust region algorithm \cite{yuan2000iciam99}. It turns out that for large sampling intervals $\tau_1$, the best results are achieved, when only that single $\tau_1$ is used, i.\,e. $M=1$ in Eq. \eqref{eq:pot}. For smaller sampling intervals the accuracy can be improved by the use of more $\tau$ values.

After the optimization procedure has converged to a certain set of parameters $\sigma_{\text res}$, one can perform a self-consistency check by comparing graphically the functions $\KM{1,2}_{\tau}(x,\sigma_{\text res})$ and $\hat D^{(1,2)}_{\tau}(x)$ as in Fig. \ref{fig:3d}.

\section{Numerical examples} \label{sec:exmpl}

\begin{figure}[t!]
  \small{\input{OU_tau1.tex}} 
  \caption{(Color online) Results for an Ornstein-Uhlenbeck process with $\KM 1(x) = -x$, $\KM 2(x) = 1$. The analyzed time series consists of $10^7$ data points with a sampling interval $\tau_1= 1$. The blue crosses with error bars are the estimated finite time coefficients $\KM 1_{\tau_1}$ (top) and $\KM 2_{\tau_1}$ (bottom). The blue dotted curves show the initial guesses for the optimization, the red solid ones show the result. For comparison also the true coefficients are plotted (black dots).} \label{fig:OU}
\end{figure}

\subsection{Ornstein-Uhlenbeck process}

As a first numerical example we consider an Ornstein-Uhlenbeck process with $\KM 1(x) = -x$ and $\KM 2(x) = 1$. A synthetic time series with $10^7$ data points is computed using a forward Euler scheme with a time step $\Delta t=10^{-3}$, but only every 1000th time step is stored. So the minimal time increment, that is available for the data analysis, is $\tau_1 = 1$. The symbols with the error bars in Fig. \ref{fig:OU} show the estimated finite time coefficients $\KM 1_{\tau_1}(x)$ (top) and $\KM 2_{\tau_1}(x)$ (bottom). From this it seems reasonable to make the parametric ansatz $\KM 1(x) = -ax$ and $\KM 2(x) = b + cx^2$. As an initial guess, we choose $a_{\text{ini}}=0.63$, $b_{\text{ini}}=0.43$ and $c_{\text{ini}}=0.2$. The corresponding curves are shown in blue in Fig. \ref{fig:OU}. 
The resulting parameters from the optimization are $a_{\text{res}}=0.9966$, $b_{\text{res}}=0.9995$ and $c_{\text{res}}=0.00032$. These values correspond to the red curves in Fig. \ref{fig:OU}. For comparison we also plot the black dots which correspond to the true parameters $a=1$, $b=1$ and $c=0$.

\subsection{Multiplicative noise} \label{sec:mul}

The next example is a system with multiplicative noise, i.\,e., the diffusion term depends on $x$. We choose $\KM 1(x) = -x$ and $\KM 2(x) = 1+x^2$. In the same manner as in the previous example, we construct a time series with $10^8$ data points and a sampling interval $\tau_1=1$. From the estimated finite time coefficients for $\KM{1,2}_{\tau_1}$ we again deduce the parametric ansatz $\KM 1(x) = -ax$ and $\KM 2(x) = b + cx^2$ and take as an initial guess $a_{\text{ini}}=0.63$, $b_{\text{ini}}=1.0$ and $c_{\text{ini}}=0.7$. The finite time coefficients as well as the initial condition are depicted in blue in Fig. \eqref{fig:mul}. From the optimization we obtain the parameters $a_{\text{res}}=0.9989$, $b_{\text{res}}=1.004$ and $c_{\text{res}}=0.9963$. The corresponding curves are shown in red in Fig. \eqref{fig:mul} as well as the true coefficients with $a=b=c=1$ (black dots).

\begin{figure}[bht]
  \small{\input{mul_noise.tex}} 
  \caption{(Color online) System with multiplicative noise: $\KM 1(x) = -x$, $\KM 2(x) = 1+x^2$. The analyzed time series consists of $10^8$ data points with a sampling interval $\tau_1= 1$. The representation is analog to Fig. \ref{fig:OU}.} \label{fig:mul}
\end{figure}

\subsection{Bistable system} \label{sec:bistable}

\begin{figure}[htb]
  \small{\input{x3.tex}} 
  \caption{(Color online) Bistable system with $\KM 1(x) = x-x^3$, $\KM 2(x) = 1$. The analyzed time series consists of $10^7$ data points with a sampling interval $\tau_1= 0.1$. The blue and red symbols are sampling points, from which the corresponding curves are computed as spline interpolations, and serve as optimization parameters. The blue dots represent the initial condition derived from the finite time coefficients $\KM{1,2}_{\tau_1}$. The red squares show the result of the optimization. The black dashed curves show the true coefficients for comparison.} \label{fig:x3}
\end{figure}

The first example for a parameter free optimization is a bistable system with $\KM 1(x) = x-x^3$ and $\KM 2(x) = 1$. The blue dots in Fig. \eqref{fig:x3} correspond to the finite time coefficients $\KM{1,2}_{\tau_1}$. They are used as the initial guess for the parameters $\sigma$ to be optimized.
The terms $M^{(1,2)}_{\tau_i}(x_j,\sigma)$ in Eq. \eqref{eq:pot} are now calculated by a spline interpolation between these sampling points.
They are shown in blue in Fig. \eqref{fig:x3}. The resulting parameters that minimize \eqref{eq:pot} are
the red squares from which the red spline curves are calculated. The latter represent the resulting drift and diffusion coefficients.

\begin{figure}[thb]
  \small{\input{phi.tex}} 
  \caption{(Color online) Phase dynamics with $\KM 1(x) = 0.2 + \cos(x)$, $\KM 2(x) = 0.5$. The analyzed time series consists of $10^7$ data points with a sampling interval $\tau_1= 1$. The representation is analog to Fig. \ref{fig:x3}}
  \label{fig:phi}
\end{figure}

\subsection{Phase dynamics}

As a last example, we consider a phase variable $\phi$ that can also be a phase difference $\phi=\phi_1 - \phi_2$ between two coupled nonlinear oscillators. The reconstruction of phase dynamics from data sets is an important theoretical problem that is relevant in many different fields of science. The problem was among others tackled by Kralemann \emph{et al.} \cite{kralemann2008pre}. We suggest the KM approach as a less cumbersome alternative.

In the case of phase dynamics, the drift and diffusion coefficients must be $2\pi$-periodic, i.\,e., $\KM n(x) = \KM n(x+2\pi)$. Therefore, it makes sense to define the KM coefficients as
\begin{equation}
  \KM n (x,t) = \lim_{\tau \rightarrow 0} \frac{1}{n! \tau} \l.\l \langle [\phi(t+\tau)- \phi(t) ]^n  \r \rangle \r|_{\phi(t)\, \text{mod}\,2\pi = x}~.
\end{equation}
Phase dynamics are often governed by Langevin equations of the form
\begin{equation}
  \dot \phi = \omega + \cos(\phi) + \sqrt{2D}\Gamma~.
\end{equation}
We consider the case $\omega=0.2\,;~D=0.5$, so we have $\KM 1(x) = 0.2 + \cos(x)$ and $\KM 2(x) = 0.5$\,. Fig. \ref{fig:phi} shows the result in the same representation as in Fig. \ref{fig:x3}.

\section{Summary and outlook}

We have presented a novel optimization procedure for the estimation of drift and diffusion coefficients for one-dimensional Markovian time series that suffer from large sampling intervals. The optimization can be performed both in a parametric and non-parametric fashion. Therefore, it is applicable for a large class of diffusion processes. The usefulness of our method is demonstrated in four examples with synthetic time series. The method yields good results, even if the sampling interval is of the order of the typical time scales of the deterministic part of the dynamics. 

\bibliography{Lit}

\end{document}